\documentclass[
 ,draft            
,numberedheadings 
  ]
  {aipproc}

\layoutstyle{6x9}

\usepackage[usenames]{color} 

\begin{document}

\title{Hydrodynamics from dynamical non-equilibrium MD.}


\author{Sergio Orlandini}{
  address={Dipartimento di Chimica, Universit\`a  ``Sapienza'', P.le A. Moro 5, 00185 Roma, Italy},
  altaddress={Consorzio Interuniversitario per le Applicazioni di Supercalcolo Per Universit\`a e Ricerca (CASPUR), Via dei Tizii 6, 00185 Roma, Italy}
}

\author{Simone Meloni$^{1,}$  
 \footnotetext[1]{To whom correspondence should be addressed: s.meloni@caspur.it} }{address={School of Physics, Room 302 UCD-EMSC, University College Dublin, Belfield, Dublin 4, Ireland},
   altaddress={Consorzio Interuniversitario per le Applicazioni di Supercalcolo Per Universit\`a e Ricerca (CASPUR), Via dei Tizii 6, 00185 Roma, Italy}}

\author{Giovanni Ciccotti}{
  address={School of Physics, Room 302B UCD-EMSC, University College Dublin, Belfield, Dublin 4, Ireland}
  ,altaddress={Dipartimento di Fisica, Universit\`a ``La Sapienza'' and CNISM, Piazzale Aldo Moro 5, 00185 Roma, Italy}
}

\begin{abstract}

{We review a dynamical approach to non-equilibrium MD (D-NEMD). We show how, using a proper simulation setup, is possible to treat interesting cases in which the initial condition is a stationary non-equilibrium state produced by a suitable dynamical system.} We then extend the class of non-equilibrium phenomena that can be studied by atomistic simulations to the case of complex initial conditions consisting in assigning a macroscopic value of a scalar or vector observable or a field. We illustrate the functioning of this method by applying it to the relaxation of an interface between two immiscible liquids. We have shown that our method generate unbiased results while this might not be the case for the often used short time average approach.
\end{abstract}

\maketitle


\section{Introduction}
By time averages, the macroscopic properties, that are the object of equilibrium statistical mechanics, emerge from the microscopic interactions among the elementary constituents of a macroscopic body in a natural way. This is the Boltzmann point of view put in practice by  molecular dynamics (MD) simulations. The situation is much less settled in the case of non-equilibrium statistical mechanics, except for the linear response to an external field \cite{Alder1970, Wainwright1971} and for stationary non-equilibrium situations \cite{Less1972, Gosling1973, Ashurst1973,hoover1973} (where the use of time averages is still justified). In the first case, the Kubo formula reduces the determination of the linear response to the calculation of equilibrium time correlation functions, which are easy to sample accurately by MD simulations; in the second case, time averages can still be used to replace ensemble averages over the unknown ensemble corresponding to the steady state (provide the system remains ergodic). Thus, for both cases, with a proper setup the physical properties of the system can be computed.

A MD method to compute the statistical properties of a non-equilibrium non-stationary systems has been proposed quite some time ago \cite{Giovanni1975}. with this method rigorous ensemble averages can be obtained by following the so called {\itshape dynamical approach} to non-equilibrium molecular dynamics (D-NEMD) \cite{Giovanni1975, jStatPhys-21_1, Ciccotti1992}. This method, which is just the generalization and the numerical implementation of the Onsager principle of regressive fluctuations \cite{PhysRev.37.405, Kubo1957}, tells that the time dependent statistical average of a microscopic observable is computed by taking an average over the initial ensemble of the observable evolved in time under a perturbed dynamics. When the initial ensemble corresponds either to a system in equilibrium or to the stationary non-equilibrium state of a given perturbation, it can be easily sampled by MD. The Onsager-Kubo relation (see below) is thus exploited to perform non-equilibrium MD ensemble averages. That involves choosing independent configurations of the system in a steady state, performing the perturbed time evolution of every
independent initial configuration and computing averages over the set of these different trajectories \cite{Giovanni1975, jStatPhys-21_1, Ciccotti1992}.
However, when the initial distribution corresponds to the equilibrium distribution of a system submitted to a macroscopic constraint, the prescription above must be complemented by a method which allows to sample the conditional Probability Density Function (PDF) associated with this condition. 
{In this paper we shall illustrate how to deal with both cases: i) initial conditions corresponding to a stationary non-equilibrium distribution \cite{Mugnai2010}, and ii) initial conditions corresponding to the conditional equilibrium distribution of a system subject to an external macroscopic constraint, be it of scalar, vectorial or field nature \cite{Orlandini201x}. This latter method allows to treat constraint of both scalar or field nature.} We will achieve this objective by combining D-NEMD with restrained MD. The latter is used to perform the conditional average satisfying the macroscopic constraint mentioned above then, like in standard D-NEMD, the Onsager-Kubo relation is used to compute time-dependent statistical average of the relaxation from this state. {It is worth to mention that the scalar and vectorial  cases have been already implicitly solved by the Blue Moon approach in Ref. \cite{Carter1989}. However, using the Blue Moon approach for vector or, even worse, field-like constraints, especially in molecular systems in which constraints are also used for imposing the molecular geometry, might result inconvenient in practice.} 

{As an illustration of the case of stationary non-equilibrium initial conditions we will study the transient hydrodynamical behavior in the formation of convective cells within a two-dimensional fluid system subject to orthogonal thermal gradient and gravity field. We follow the time-dependence of the density $\rho(x;t)$, velocity ${\textbf v}(x;t)$ and temperature $T(x;t)$ fields induced adding a gravity to a system in stationary conditions under the thermal  gradient.}

{
To illustrate our method to sample from a microscopic system under macroscopic constraints,} we study the hydrodynamic relaxation to the equilibrium of the interface between two immiscible liquids. We compute the time evolution of the difference of density fields of the species A and B ($\Delta \rho(x;t) = \rho^A(x;t) - \rho^B(x;t)$) and the associated velocity field ($v^A(x;t)$) and show the usefulness of our approach that does not rely on the separation of timescale of atomistic and hydrodynamical processes, as it is the case for the often used method of local time average.

The paper is organized as follows. In Sec. \ref{Sec:Theory} we describe the theoretical background of D-NEMD and our method for sampling complex initial conditions. {In Sec. \ref{Sec:ConvectiveCells} we present an application of D-NEMD to the study of the formation of convective cells in a system subject to thermal gradient and gravity field. In Sec. \ref{Sec:Interface} we discuss an application of our combined restrained MD + D-NEMD method.} Finally, in Sec. \ref{Sec:Conclusions} we draw some conclusions.

\section{Theoretical background}
\label{Sec:Theory}
In D-NEMD we consider a classical system with $N$ particles in $d$ spatial dimensions. Let ${\textbf r}_i$ and ${\textbf p}_i$ be the position and the momentum of the {\itshape i}-th particle, respectively, and $\Gamma = \{{\textbf r}_i , {\textbf p}_i\}$ be a point in phase space. 
The Hamiltonian governing the system is $H(\Gamma, t)$, which we assume to be explicitly time dependent (the generalization to non-Hamiltonian systems could be easily worked out \cite{Tuckerman1999,Tuckerman2001,Hartmann2010}).

In statistical mechanics an observable, 
including a macroscopic field (see below for its definition), is obtained as an ensemble average in phase space of the corresponding microscopic observable:

\begin{eqnarray}
\label{eq:averageValueObservable}
O(t) = \int d\Gamma {\hat O}(\Gamma) w(\Gamma, t) \equiv \left < {\hat O}(\Gamma) w(\Gamma, t) \right >
\end{eqnarray}

\noindent where $O(t)$ is the macroscopic observable and ${\hat O}(\Gamma)$ is the corresponding microscopic observable. $w(\Gamma, t)$ is the (time dependent) phase-space PDF. As a consequence of the conservation of probability, $w(\Gamma, t)$  obeys the Liouville equation:

\begin{eqnarray}
\label{eq:Liouvile}
{ \partial w(\Gamma, t) \over \partial t} &=& - \nabla_\Gamma [{\dot \Gamma} w(\Gamma, t)] \\
&& \{ w(\Gamma, t), H(\Gamma, t)\} = -iL(t)w(\Gamma, t) \nonumber 
\end{eqnarray}

\noindent where $\{\cdot,\cdot\}$ is the Poisson bracket and $L(t)$ is the Liouville operator. 
A formal solution of Eq.(\ref{eq:Liouvile}) is:

\begin{eqnarray}
\label{eq:timeEvolutionPDF}
w(\Gamma, t)  &=& S^\dag(t) w(\Gamma, 0)
\end{eqnarray}

\noindent where $S^\dag(t)$ is the adjoint time evolution operator of the dynamical system. 
%

We now consider the time-evolution of a non-time-dependent phase-space observable:

\begin{eqnarray}
\label{eq:TimeDerivativeObservable}
{ d  {\hat O}(\Gamma) \over d t} &=& \{  {\hat O}(\Gamma), H(\Gamma, t)\} = iL(t) {\hat O}(\Gamma) 
\end{eqnarray}

\noindent which has the following formal solution

\begin{eqnarray}
\label{eq:timeEvolutionOperator}
 {\hat O}(\Gamma(t))  &=& S(t)  {\hat O}(\Gamma(0))
\end{eqnarray}

\noindent where $S(t)$ is the time-evolution operator.

By combining Eq.(\ref{eq:averageValueObservable}) with Eq. (\ref{eq:timeEvolutionPDF}) and (\ref{eq:timeEvolutionOperator}) we get the Onsager-Kubo relation:

\begin{eqnarray}
\label{eq:Onsager-Kubo}
O(t) &=& <{\hat O}(\Gamma) S^\dag(t) w(\Gamma,0)> \\
&&<S(t) {\hat O}(\Gamma)  w(\Gamma,0)> \nonumber
\end{eqnarray}

The meaning of the above equation is that the ensemble average of the microscopic observable ${\hat O}(\Gamma)$ over the time-dependent PDF $w(\Gamma,t)$ at time $t$ is the same as the ensemble average of the microscopic observable at the point $\Gamma(t)$, corresponding to the evolution in time of the initial phase-space point $\Gamma(0)$, averaged over the PDF at time $t=0$ ($w(\Gamma,0)$).

In the standard D-NEMD approach we assume that $w(\Gamma,0)$ corresponds to a stationary condition, then can be sampled via an MD simulation possibly governed by the Hamiltonian $H_0(\Gamma)$ (more general stationary conditions can be constructed, see below and in Ref. \cite{Mugnai2010}). Then we can evolve initial configurations taken from that trajectory with the dynamics generated by $H(\Gamma, t)$. Along these paths we compute the microscopic observable ${\hat O}(\Gamma(t))$. The time-dependent behavior of the macroscopic observable $O(t)$ is the ensemble average of ${\hat O}(\Gamma(t))$ over all the trajectories originated from each of the initial states (see Eq. (\ref{eq:Onsager-Kubo})). { An example of how to generate stationary non-equilibrium conditions might be by putting the sample in contact with two thermal reservoirs at different temperature at the border of the simulation box. Using this setup it is possible by MD to sample an initial stationary PDF corresponding to a system under thermal gradient.}

We will now generalize the standard D-NEMD approach to the case of an initial PDF associated with a macroscopic constraint, i.e. the conditional PDF to find the system at the point $\Gamma$ in phase space given that a microscopic field ${\hat F}({\textbf x}| \textbf r)$, function of the atomic position only, at the time $t=0$ is equal to $F^*({\textbf x})$ (${\hat F}({\textbf x}| \textbf r) = F^*({\textbf x})$). Our method a fortiori applies to the case in which the observable $F$ is a scalar. For sake of clarity we first introduce our method assuming that $F$ is scalar and then explain how to extend it to the case of space-dependent fields. Let us start recalling that the conditional PDF mentioned above reads:

\begin{eqnarray}
\label{eq:condProbabilityDensity}
w(\Gamma | F = F^*) = {w(\Gamma) \delta({\hat F}({\textbf r}) - F^*) \over {\mathcal Z} P_F( F^*)}
\end{eqnarray}

\noindent where ${\hat F}({\textbf r})$ is the microscopic observable associated with the macroscopic observable $F$, $\mathcal Z$ is the partition function associated with $w(\Gamma)$ and $P_F( F^*) = \int d\Gamma w(\Gamma) \delta({\hat F}({\textbf r}) - F^*) / \mathcal Z$ is the PDF to find the system in the state $F =  F^*$. 

We propose to sample this conditional PDF by the biased MD simulation governed by the following Hamiltonian:

\begin{eqnarray}
\label{eq:biasedH}
{\tilde H}(\Gamma) = H_0(\Gamma) + {k \over 2} ({\hat F}({\textbf r}) - F^*)^2
\end{eqnarray}

\noindent where  $k$ is a tunable parameter. The PDF sampled by this biased MD is:

\begin{eqnarray}
\label{eq:biasedPDF}
{w}_k(\Gamma) &=& {\exp [ -\beta {\tilde H}(\Gamma)]  \over \int d\Gamma \exp [ -\beta {\tilde H}(\Gamma)] } \\
&=& {\exp [ -\beta (H_0(\Gamma) + {k \over 2} ({\hat F}(\textbf r) - F^*)^2) ]  \over \int d\Gamma \exp [ -\beta (H(\Gamma, 0) + {k \over 2} ({\hat F}(\textbf r) - F^*)^2)] } \nonumber
\end{eqnarray}

By recalling that  $\exp(-(y - \mu)^2 / 2 \sigma^2) / \sqrt{2 \pi} \sigma \rightarrow \delta(y - \mu)$, in the limit of $k \rightarrow \infty$, we see that Eq. (\ref{eq:biasedPDF}) goes to

\begin{eqnarray}
\label{eq:biasedPDFLimit}
{w}_k(\Gamma) \rightarrow  {\exp [ -\beta (H_0(\Gamma)] \delta({\hat F}(\textbf r) - F^*)  \over \int d\Gamma \exp [ -\beta (H_0(\Gamma)] \delta({\hat F}(\textbf r) - F^*) }
\end{eqnarray}

\noindent Multiplying and dividing by $\mathcal Z$ it is apparent that ${w}_k(\Gamma)  \rightarrow w(\Gamma | F = F^*)$.

We now extend our discussion to the case in which the condition (and the observables of interest) is with respect to a macroscopic field, rather than to a scalar. Generally speaking, in statistical mechanics the microscopic observable associated to a macroscopic field is defined as \cite{irving:817}:

\begin{eqnarray}
\label{eq:field}
{\hat F}({\textbf x}, \Gamma) = \sum_{i=1}^N {\mathcal F}_i(\Gamma) \delta({\textbf x} -  {\textbf r}_i)
\end{eqnarray}

\noindent where $ {\mathcal F}_i(\Gamma)$ is the microscopic property under consideration relative to the particle $i$ and ${\textbf x}$ is a point in the ordinary $\Re^3$ space. Eq. (\ref{eq:field}) means that only atoms at the point ${\textbf r}_i = {\textbf x}$ contribute to the field at that point. Examples of fields of interest for the problem presented in Sec. \ref{Sec:ConvectiveCells} and  \ref{Sec:Interface} are the fields density ($\rho({\textbf x}, t)$), velocity (${\textbf v}({\textbf x}, t)$) and temperature ($T({\textbf x}, t)$). These fields at time $t$ are given by:

\begin{eqnarray}
\label{eq:fieldDensity}
\rho({\textbf x}, t) = <\sum_{i=1}^N \mu_i \delta({\textbf x} -  {\textbf r}_i) w(\Gamma, t)>
\end{eqnarray}

\begin{eqnarray}
\label{eq:fieldVelocity}
{\textbf v}({\textbf x}, t) = {<\sum_{i=1}^N {\textbf p}_i \delta({\textbf x} -  {\textbf r}_i) w(\Gamma, t)> \over \rho({\textbf x}, t) }
\end{eqnarray}

\begin{eqnarray}
\label{eq:fieldTemperature}
T({\textbf x}, t) = {1 \over 3 k_B}{<\sum_{i=1}^N  \delta({\textbf x} -  {\textbf r}_i) [{\textbf p}_i - \mu_i {\textbf v}({\textbf x},t)]^2 w(\Gamma, t)> \over \rho({\textbf x}, t) }
\end{eqnarray}

\noindent where $\mu_i$  is the mass of the $i$-th atom. 

Eq. (\ref{eq:condProbabilityDensity}) can be easily extended to the case in which the condition is a field:

\begin{eqnarray}
\label{eq:condProbabilityDensityField}
w[\Gamma | {\hat F}({\textbf x}, \Gamma) = F^*({\textbf x})] = {w(\Gamma) \delta({\hat F}({\textbf x}, \Gamma) - F^*({\textbf x})) \over {\mathcal Z} P_F[F^*({\textbf x})]}
\end{eqnarray}

\noindent Now $w[\Gamma | {\hat F}({\textbf x})]$ is functional of the field $F({\textbf x})$ and the notation $\mathbf{\delta}({\hat F}({\textbf x}| \Gamma) - F^*({\textbf x})) \equiv \prod_{{\textbf x} \in \Re^3} \delta({\hat F}({\textbf x}, \Gamma) - F^*({\textbf x}))$ indicates that the delta function is considered acting all over the $\textbf x$ space.
Of course, this definition of the conditional probability density functional is of no practical use in simulations. We overcome this problem by introducing a discretization $\{{\textbf x}_\alpha\}_{\alpha=1,m}$ of the $\Re^3$ space, where $m$ is the number of points 
over which the space is discretized. Consistently with this discretization the microscopic observable fields is defined as the average over the cells around around each point ${\textbf x}_\alpha$:

\begin{eqnarray}
\label{eq:field-Discrete}
{\hat F}({\textbf x}_\alpha, \Gamma) &=& {1 \over \Omega_\alpha} \int_{{\Omega}_\alpha} d{\textbf x} \sum_{i=1}^N {\mathcal F}_i(\Gamma) \delta({\textbf x} -  {\textbf r}_i) =  {1 \over \Omega_\alpha} \sum_{l=1}^{N_\alpha} {\mathcal F}_l(\Gamma)
\end{eqnarray}

\noindent where the sum in the r.h.s. runs over the atoms belonging to the cell around the point ${\textbf x}_\alpha$. However, this definition is not suitable for our restrained MD method as it might produce impulsive forces on the atoms moving from one cell to another (see below). Therefore we smooth the central term of Eq. (\ref{eq:field-Discrete}) by replacing  $\delta({\textbf x} -  {\textbf r}_i)$ with the (normalized) gaussian $g({\textbf x}; {\textbf r}_i, \sigma) = \exp[-({\textbf x} -  {\textbf r}_i)^2/2 \sigma^2]/\sqrt{2\pi}\sigma$. With this approximation, Eq. (\ref{eq:field-Discrete}) becomes:

\begin{eqnarray}
\label{eq:field-Discrete-Smooth}
{\hat F}({\textbf x}_\alpha, \Gamma)&=&\\
 {1 \over \Omega_\alpha}&\sum_{i=1}^{N}& {\mathcal F}_i(\Gamma) \Pi_{\chi=1}^3 \left [ erf({\overline x}_{\alpha,\chi} -  r_{i,\chi}, \sigma) - erf({\underline x}_{\alpha,\chi} -  r_{i,\chi}, \sigma) \right ] \nonumber
\end{eqnarray}

\noindent where $erf(c, \sigma) = \int_{-\infty}^c dx\ g(x; 0, \sigma)$ is the error function, and ${\overline x}_{\alpha,\chi}$ and ${\underline x}_{\alpha,\chi}$ is the $\chi$ component of the upper and lower limit of the orthorombic cell around the point ${\textbf x}_\alpha$. 

The conditional PDF associated to the field $F({\textbf x})$ on the discrete representation of the $\Re^3$ space is:

\begin{eqnarray}
\label{eq:condProbabilityDensityField-Discrete}
w\left (\Gamma | \{{\hat F}({\textbf x}_\alpha, \Gamma) = F^*({\textbf x}_\alpha)\}_{\alpha=1,m} \right) = {w(\Gamma) \Pi_{\alpha=1}^m \delta({\hat F}({\textbf x}_\alpha, \Gamma) - F^*({\textbf x}_\alpha)) \over {\mathcal Z} P_F(\{F^*({\textbf x}_\alpha)\}_{\alpha=1,m})}
\end{eqnarray}

\noindent where $P_F(\{F^*({\textbf x}_\alpha)\}_{\alpha=1,m})$ is the joint probability that ${\hat F}({\textbf x}_1, \Gamma) = F^*({\textbf x}_1)$ $, \dots,$ ${\hat F}({\textbf x}_m, \Gamma) = F^*({\textbf x}_m)$. This conditional PDF can be sampled by a MD governed by the following Hamiltonian

\begin{eqnarray}
\label{eq:biasedH-field}
{\tilde H}(\Gamma) = H_0(\Gamma) + \sum_{\alpha = 1}^m{k \over 2} ({\hat F}({\textbf x}_\alpha, {\textbf r}) - F^*({\textbf x}_\alpha))^2
\end{eqnarray}

\noindent which is the straightforward extension of the Hamiltonian of Eq. (\ref{eq:biasedH-field}) to the case of a macroscopic field in the discrete space approximation. Then, as in standard D-NEMD approach, we evolve a set of initial configurations taken from the trajectory above with the dynamics generated by $H(\Gamma, t)$. Along these paths we can compute the microscopic observables and calculate the  ensemble average over all the trajectories originated from each initial state (Eq. (\ref{eq:Onsager-Kubo})). 

{Before closing this section it is worth to mention that the problem of sampling initial conditions consistent with a macroscopic constraint was already implicitly solved by the Blue Moon ensemble (see Refs. \cite{Carter1989} and \cite{VandenEijnden2005}). Blue Moon does not sample directly the conditional PDF $w(\Gamma | F = F^*)$ (or $w\left (\Gamma | \{{\hat F}({\textbf x}_\alpha, \Gamma) = F^*({\textbf x}_\alpha)\}_{\alpha=1,m} \right)$ for the vectorial case) but rather it samples the constrained PDF in configurational space $w_{F^*}({\textbf r})$. The relation between the Blue Moon ensemble average and  conditional average in the configurational space is given by (see Ref. \cite{VandenEijnden2005}):}

\begin{eqnarray}
\label{eq:ConditionalVsBMSampling}
<{\hat O}(\textbf{r})>_{F = F^*} = { <{\hat O}(\textbf{r}) 1 / \sqrt{|\det[C({\textbf{r}})]|} >_{BM} \over <1 / \sqrt{|\det[C({\textbf{r}})]|} >_{BM} } 
\end{eqnarray}

\noindent  {where ${\hat O}(\textbf{r})$ is a microscopic observable and $C_{ij}({\textbf{r}}) = {\mathbf \nabla} \sigma_i({\textbf{r}}) M^{-1} {\mathbf \nabla} \sigma_j({\textbf{r}})$, being $\sigma_i({\textbf{r}})$ the (vectorial) condition ${\hat F}({\textbf{r}}) = F^*$ or any other constraint imposed on the system, in particular constraints imposed for modeling  (partly) rigid molecules \cite{Ciccotti1986}, and $M$ the mass matrix ($M_{ij} = \mu_i \delta_{ij}$). The $F = F^*$ and $BM$ indexes denote that the ensemble averages are taken according to the conditional or constrained (Blue Moon) PDF, respectively. If the observable of interest depends on the phase space (${\hat O}(\Gamma)$) rather than on the configurational space it is possible to extend the validity of Eq. \ref{eq:ConditionalVsBMSampling} by generating the momentum component of the PDF from a suitable Maxwellian distribution.}

Two comments are in order concerning the restraint method for sampling the complex initial condition described in this paper versus the approach based on the Blue Moon ensemble. First of all, depending on the macroscopic condition and the molecular constraints, calculating the unbiasing term $1 / \sqrt{|\det[C({\textbf{r}})]|} / <1 / \sqrt{|\det[C({\textbf{r}})]|} >_{BM}$ might be complex. Moreover, while the restraint approach can be combined with Monte Carlo simulation when the microscopic observable connected to the macroscopic condition is not analytical (see Ref. \cite{Meloni201x} for a detailed description of this approach), the same cannot be done with the Blue Moon approach.

In the following two sections we show two applications of the D-NEMD described above. We first present the case of an initial condition corresponding to a stationary non-equilibrium system. In particular we study the transient state of formation of convection cells when a gravity force is added to a system subject to a thermal gradient  (Sec. \ref{Sec:ConvectiveCells}). Then we illustrate an application of the restrained method to sample initial conditions corresponding to a system subject to macroscopic constraints by studying the relaxation to the equilibrium of the interface between two immiscible liquids (Sec. \ref{Sec:Interface}).

\section{Formation of convective cells}
\label{Sec:ConvectiveCells}

\subsection{Computational Model and Setup.}
\label{Sec:ConvectiveCellsModelAndSetup}
{A fluid system consisting of $N = 5041$ particles is contained in a two- dimensional box in the $\{xz\}$ plane, with the gravity force directed along the negative verse of the z axis. The particles interact via a WCA (Weeks-Chandler-Andersen) potential}

\begin{equation}
u(r) = \left \{
\begin{array}{l}
 4 \epsilon \left [ (\sigma / r)^{12} - (\sigma / r)^{6}\right ], \forall\ r \leq 2^{1/6} \sigma\\
0, \forall\ r \ge 2^{1/6} \sigma
\end{array}
\right .
\end{equation}

{\noindent where $\epsilon$ and $\sigma$ are the usual Lennard-Jones parameters. This potential is a purely repulsive potential obtained from the Lennard-Jones 12-6 potential, truncated in its minimum (so that the force is continuous), and shifted (so that the potential is continuous). Each particle has mass $\mu$. Hereafter, we will use reduced units putting the typical scale of energy, mass, and length equal to unity, i.e., $\epsilon = 1$, $\mu = 1$,	and	$\sigma = 1$. Times are then measured  in units of $\sqrt{\mu \sigma^2/\epsilon}$.}

{We want to study the system in physical conditions that
allow the formation of a convective cell, i.e., when gravity and a thermal gradient are present. Moreover, we want to analyze the transient evolution to the formation of the steady-state roll. For this we can take as initial condition of the system the steady state under the effect of a thermal gradient and then study the dynamical response of the system to the ignition of gravity. As far as the physical setting is concerned, we take  the thermal gradient orthogonal to the gravity force. This allows a straightforward application of the D-NEMD technique. At the top and at the bottom of the box there are repulsive walls to avoid particles from drifting downwards under the effect of the gravity force. The thermal reservoirs which produce the thermal gradient are realized as two stripes such that the components of the velocity of each particle located in one of these stripes are sampled from a Maxwellian distribution at the temperature of the wall. We assume periodic boundary conditions at the thermal walls located at the two lateral sides of the box, i.e., a particle can move from the hotter to the colder reservoir. To avoid that particles near a thermal reservoir interact simultaneously with both reservoirs, we chose the thickness of the reservoir	$x_T = 1.68$, larger than the cutoff of the WCA interaction. When the system is in equilibrium, each reservoir contains roughly 100 particles on average. We chose the temperature of the colder reservoir as $T_1 = 1.5$ and a theoretical thermal gradient $|\nabla T| = 0.1$, so that the hotter reservoir has a temperature $T_2 = |\nabla T| L + T_1 = 9.9$. The gravity force used is $g = 0.1$ in Lennard-Jones units. In SI units, taking Ar as a reference fluid, we have $g = 7 \times 10^{12}$~m/s$^2$, $T_2=1196$~K, and $T_1=179.7$~K. The box length is $L = 2.89  \times 10^{-8}$~m, so that the thermal gradient is $|\nabla T| = (T_2 - T_1) / L = 3.52  \times 10^{10}$~K/m. External	fields of this strength	are necessary to reach a sufficient signal to noise ratio in a small system. The density, velocity and temperature fields are computed on a $15 \times 15$ discretization of the $\Re^2$ space. Finally, The averages for these simulations are performed over an ensemble made of 1000 copies of the system}

\subsection{Results and discussion.}
\label{Sec:ConvectiveCellsResultsAndDiscussion}
\begin{figure}[h]
\includegraphics[width=1\textwidth]
{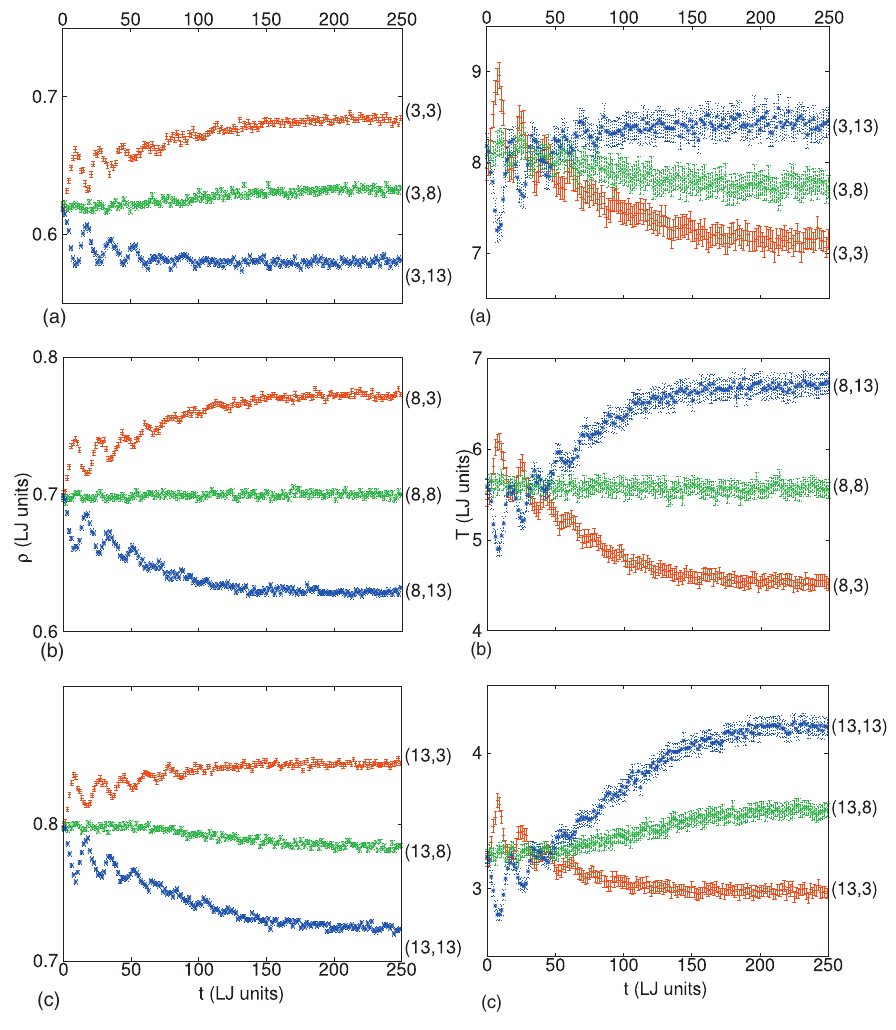}
\caption{Density (left) and temperature (right) fields as a function of time in selected cells near the hot reservoir (top panel), in the middle (middle panel) and near the cold reservoir (bottom panel). For a given distance from the reservoirs, three cells at different height are shown. 
}
\label{Fig:DensityTemperatureConvectiveCells}
\end{figure}

{We follow the evolution of the system after the igniton of the gravity field by monitoring the time evolution of the temperature and density fields in some characteristic cells (see Fig. \ref{Fig:DensityTemperatureConvectiveCells}). These fields become stationary at $t = 250$. During the transient, both temperature and density oscillate with nearly the same period $\tau = 18$ in all cells. The phase of these oscillation is constant in all cells at the bottom of the box, and it is the same for $\rho$ and $T$. The same feature holds for the phase in the cells at the top of the box, which is, however, opposite to the phase in the cells at the bottom. During the transient, an increase (decrease) in T corresponds to an increase (decrease) in $\rho$ in the same cell, while, in the stationary state at large time, a temperature lower (higher) than the initial value is associated with a density higher (lower) than the initial value, as expected.}

\begin{figure}[h]
\includegraphics[width=1\textwidth]{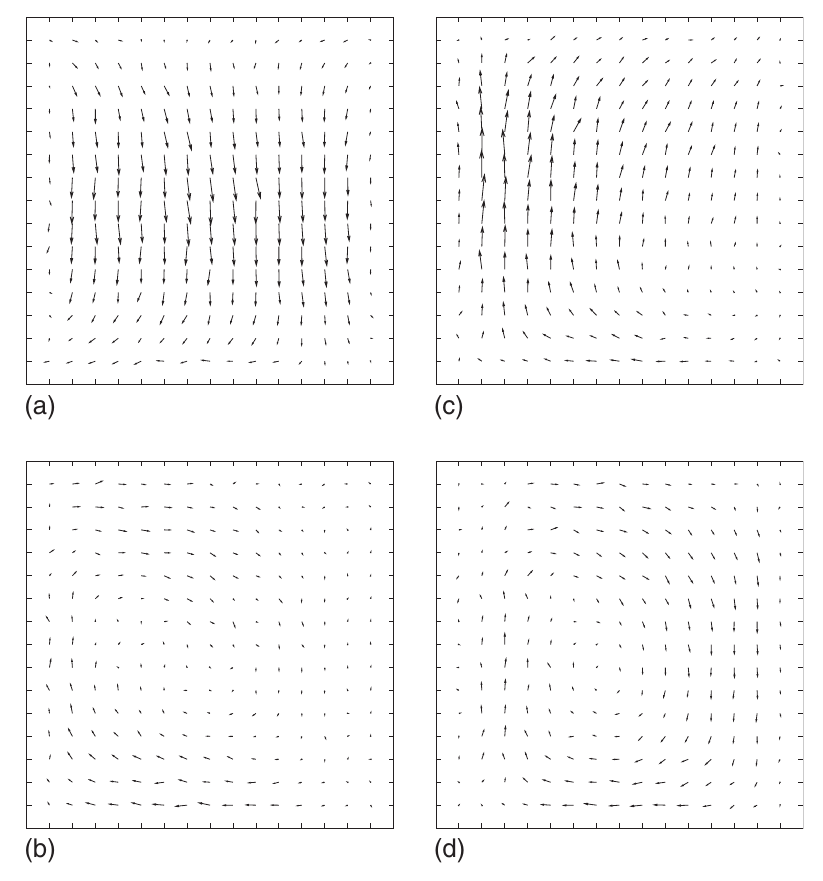}
\caption{{Local velocity field averaged over 1000 independent initial configurations during the first oscillation of temperature and density fields, i.e., at (a) $t = 4.5 \sim \tau/4$, (b) $t = 9 \sim \tau/2$, (c) $t = 13 \sim 3\tau/4$, and (d) $t = 18 \sim \tau$. Hot reservoir on the left side of the box, cold reservoir on the right side of the box.}}
\label{Fig:transientVelocityConvectiveCells}
\end{figure}

\begin{figure}[h]
\includegraphics[width=1\textwidth]{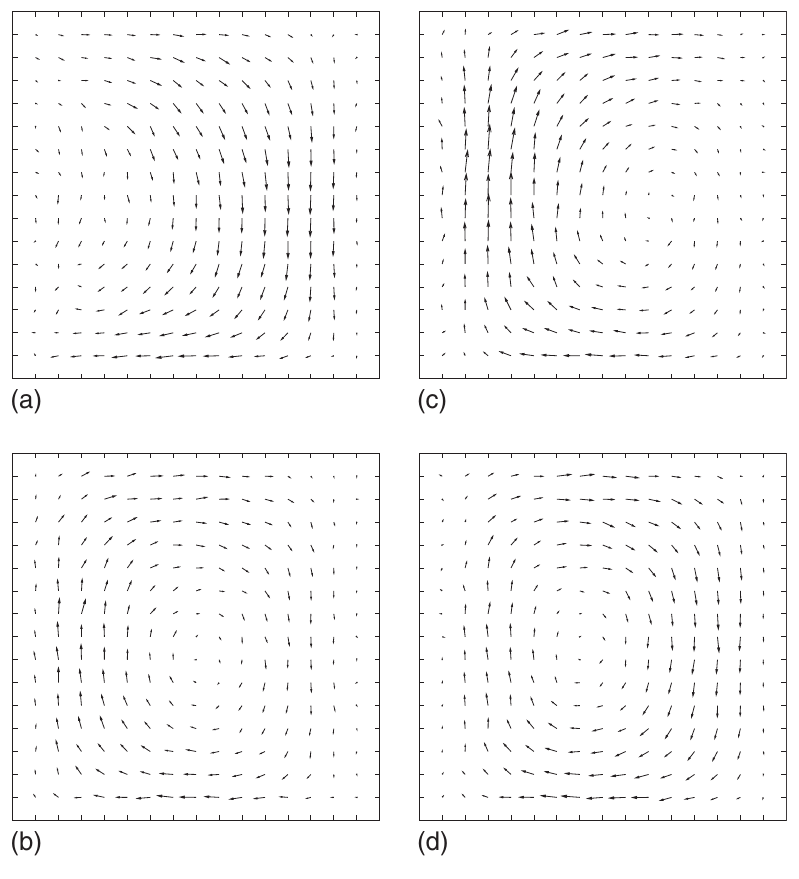}
\caption{{Local velocity field averaged over 1000 independent initial configurations during the first oscillation of temperature and density fields, i.e., at (a) $t = 22.5 \sim 5 \tau/4$, (b) $t = 27 \sim 3 \tau/2$, (c) $t = 31.5 \sim 7 \tau/4$, and (d) $t = 36 \sim 2 \tau$. Same conditions as in Fig. \ref{Fig:transientVelocityConvectiveCells}}}
\label{Fig:transient2VelocityConvectiveCells}
\end{figure}

\begin{figure}[h]
\includegraphics[width=1\textwidth]{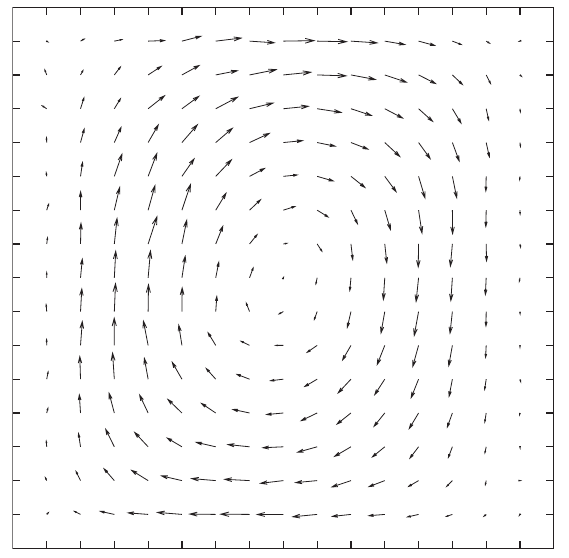}
\caption{{Local velocity field averaged over 1000 independent initial configurations at $t = 250$, i. e. at the end of simulation.  Same conditions as in Fig. \ref{Fig:transientVelocityConvectiveCells}}}
\label{Fig:stableVelocityConvectiveCells}
\end{figure}

{This behavior can be followed by plotting the transient velocity field at every quarter period of the density and temperature oscillations (Fig. \ref{Fig:transientVelocityConvectiveCells}). At	$t = \tau / 4$ (Fig. \ref{Fig:transientVelocityConvectiveCells}/a)	the velocity field points downwards as a consequence of the ignition of gravity at $t = 0$. At $t = \tau/2$ (Fig. \ref{Fig:transientVelocityConvectiveCells}/b), the velocity field is almost null, and the fluid is almost at rest on average.	 At $t = 3\tau / 4$ (Fig. \ref{Fig:transientVelocityConvectiveCells}/c) the fluid is expanding against the gravity force and the velocity field is directed upward, as a reaction to compression. Once again, in correspondence to the subsequent relative maxima or minima (Fig. \ref{Fig:transientVelocityConvectiveCells}/d) the fluid is nearly at rest. 
During the following compressions and expansions the flux becomes localized, respectively, near the cold the hot reservoirs (see Fig. \ref{Fig:transient2VelocityConvectiveCells}). In correspondence of the next relative maxima and minima the fluid is no longer at rest, and instead it begins to support a convective flow. The sequence of compressions and expansions strengthens the flux, which becomes stable at $t = 250$. The final shape of the convective pattern is symmetric (see Fig. \ref{Fig:stableVelocityConvectiveCells}).}

\section{Relaxation of the interface between two immiscible liquids}
\label{Sec:Interface}

\subsection{Computational Model and Setup.}
\label{Sec:ModelAndSetup}
We now illustrate the method described in Sec. \ref{Sec:Theory} for sampling complex initial conditions by studying the relaxation to equilibrium of the interface between two immiscible liquids. We define the interface between the two liquids as the surface at which $\Delta \rho({\textbf x}) = \rho^A({\textbf x}) - \rho^B({\textbf x}) = 0$, where $\rho^A({\textbf x})$ and $\rho^B({\textbf x})$ are the densities of the liquids A and B (see Eq. (\ref{eq:fieldDensity})), respectively. We start from the interface defined below and follow the isosurface $S(t) = \left \{{\textbf x}:  \Delta \rho({\textbf x}, t) = 0 \right \}$ of the fields $\Delta \rho({\textbf x}, t)$ and ${\textbf v}^A({\textbf x}, t)$ till equilibrium. The initial conditional PDF is sampled using the method described in Sec. \ref{Sec:Theory} with the restraint that $\Delta \rho({\textbf x}_\alpha) = 0$ in the cells, centered around the points ${\textbf x}_\alpha$,  through which passes the following surface:

\begin{equation}
S = \left \{ {\textbf x}:  x_3 =  {\mathcal A} \sin \left(  {\pi x_1 \over L_1 } \right) + {L_3 \over 2} - {A \over 2} \right \}
\end{equation}

\noindent where $\mathcal A$ is the amplitude of the curved surface and $\{L_\chi\}_{\chi=1,3}$ is the length of the simulation box along the $\chi$-th cartesian direction. The terms $ {L_3 \over 2}$ and $-{{\mathcal A} \over 2}$ are added to place the interface at the centre of the simulation box. The condition above is the discrete counterpart of the continuos condition $\Delta \rho({\textbf x}) = 0, \forall\ {\textbf x} \in S$. We do not impose any other condition on the density.  However, we prepare the system such that all the particles on one side of the interface are of one kind, say A, and of the other kind on the other side, say B. Since we apply periodic boundary conditions along all the cartesian directions, we have a second flat interface at the beginning/end of the simulation box. 

The sample used in our simulation consists of 171,500 particles: 88,889 of type A and 82,611 of type B. Two particles of the same type interact via  Lennard-Jones potential $u^{AA}(r) \equiv u^{BB}(r) = 4 \epsilon \left [ (\sigma / r)^{12} - (\sigma / r)^{6}\right ]$, while two particles of different type interact via the repulsive potential $u^{AB}(r) = 4 \epsilon \left [ (\sigma / r)^{12} \right ]$, where $\epsilon$ and $\sigma$ are the usual Lennard-Jones parameters. The simulation box is a parallelepiped of size $\sim 45 \times 45 \times 90$ in reduced units (average density $\rho = 1.024 particles/\sigma^3$). In the restrained MD the temperature of the sample is kept fixed at $1.5 \epsilon/k_B$ ($k_B$ Boltzmann constant). This density and temperature are in the fluid domain of a pure Lennard-Jones system. The ordinary space is discretized in 5488 points (a $14 \times 14 \times 28$ grid) and each cell contains, on average, $\sim 30$ particles.

\begin{figure}[h]
\includegraphics[width=1\textwidth]{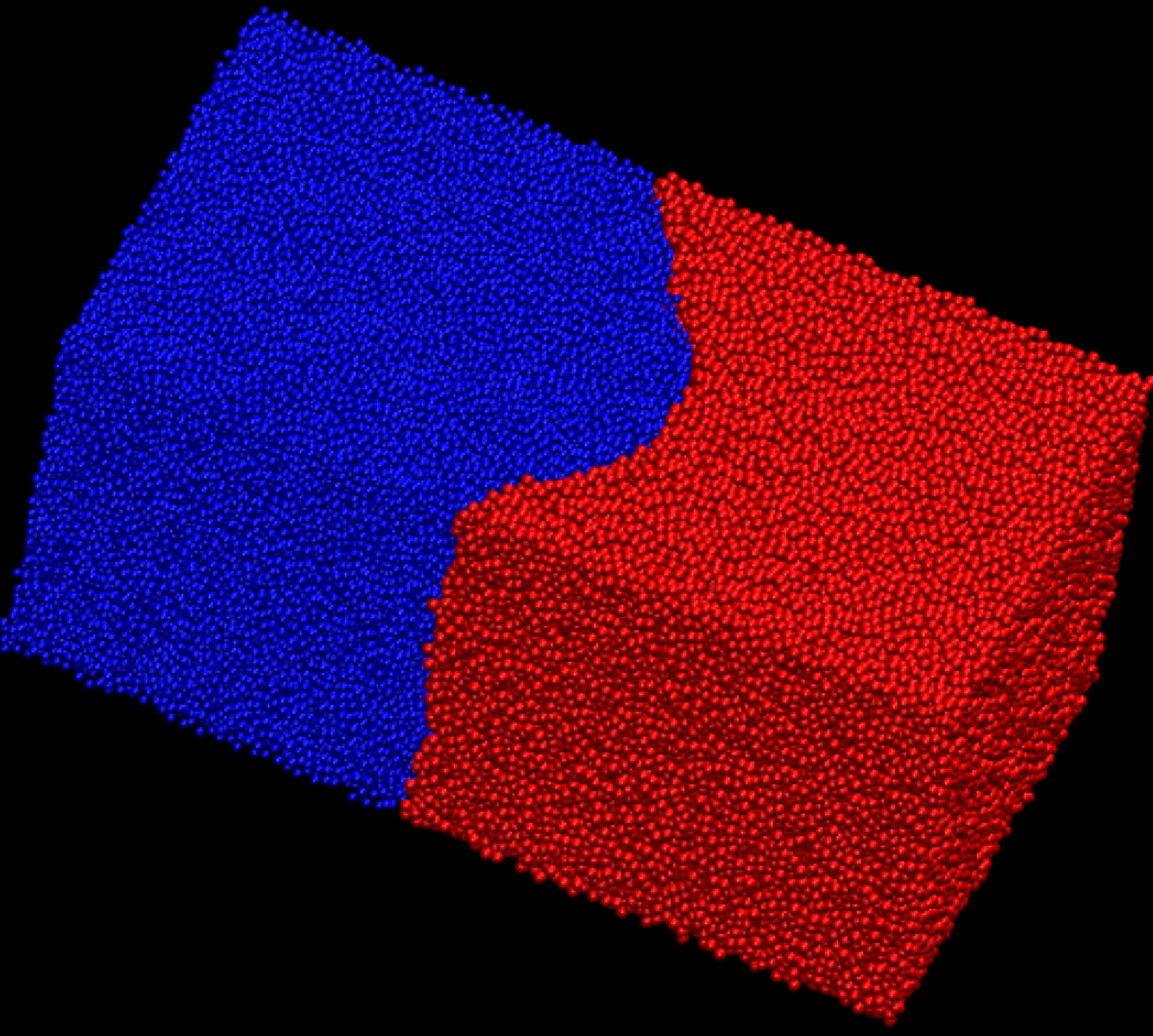}
\caption{One atomic configuration extracted from the restrained MD used for sampling the initial conditional PDF. In blue particles of type A, in red particles of type B.}
\label{fig:snapshot}
\end{figure}

The system is prepared by thermalizing a sample of pure type A Lennard-Jones particles at the target temperature and density, and then transforming those particles belonging to the cells on one side of the interface in type B particles (see Fig. (\ref{fig:snapshot})). In the cells belonging to the interface $S({\textbf x})$ only half of the particles are transformed from A to B, so as to have $\Delta \rho({\textbf x}) = 0$ in these cells. The system is then thermalized with the restraint on the $\Delta \rho({\textbf x})$ for $1.6 \times 10^6$ timesteps. Such a very long run is needed to relax the gradient of temperature formed when the nature of particles on one side of the interface is changed from A to B (immediately after the A-to-B transformation the interfaces - the curved one and the flat one due to the periodic boundary conditions - due to the strong repulsive forces among particles of different type, are warmer than the bulk). The timestep used in this and the next phase (restrained MD runs) is $4.56 \times 10^{-4}$ LJ time units, which is one order of magnitude smaller than the typical timestep for simulation of Lennard-Jones systems. This very short timestep is required by the stiff force associated to the restraint. After this relaxation, a $10^6$ timestep long restrained MD is performed along which, at regular intervals of 25,000 timesteps, we collect 40 initial positions and velocities for the second step of the D-NEMD procedure. The atomic configuration corresponding to one snapshot of this trajectory is shown in Fig. (\ref{fig:snapshot}). 

\begin{figure}[h]
\includegraphics[width=1\textwidth]{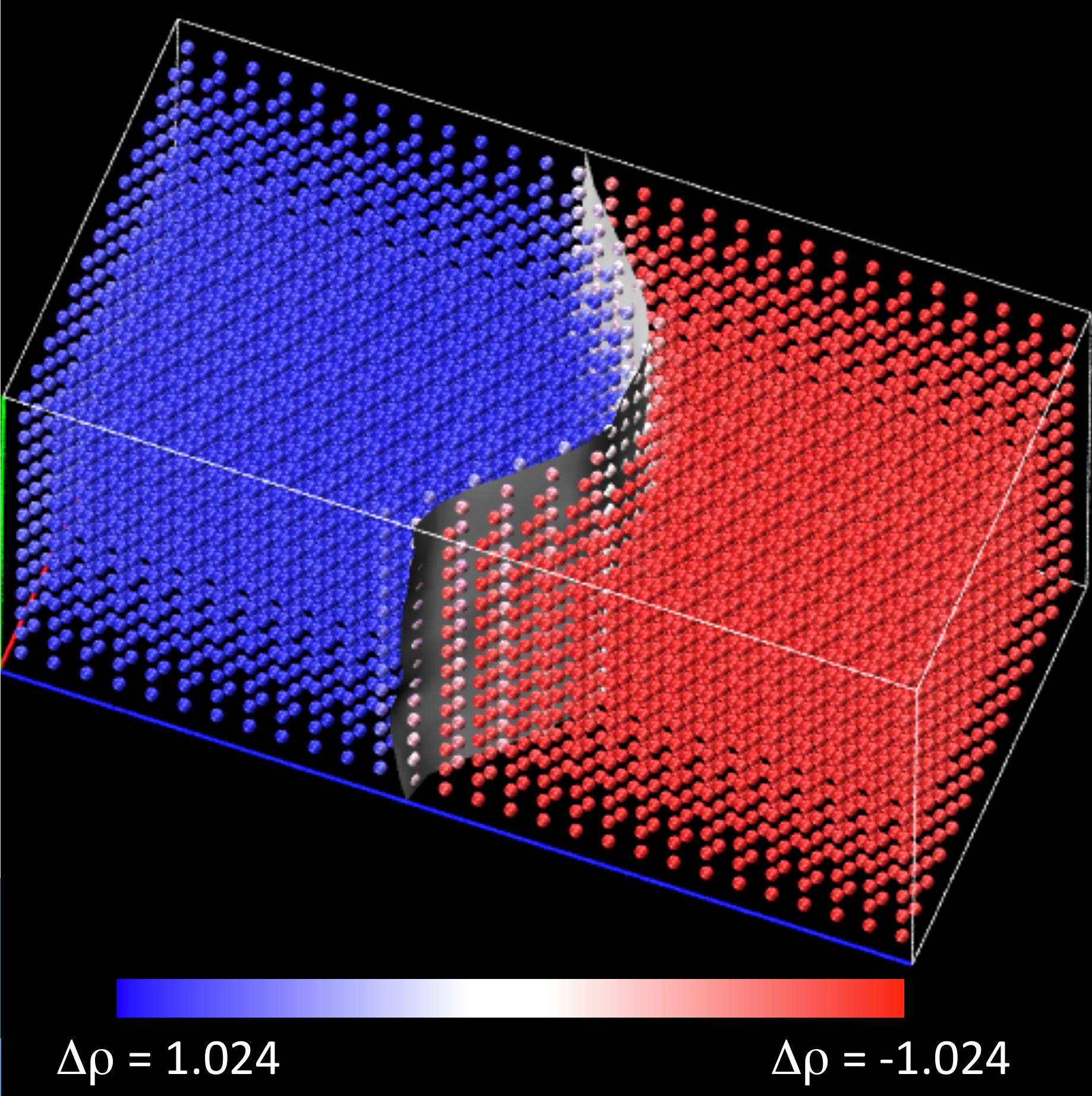}
\caption{$\Delta \rho({\textbf x})$ field over the grid points $\{{\textbf x_\alpha}\}_{\alpha=1,m}$. The color of each point depends on the $\Delta \rho({\textbf x})$ on that point. Intense red means that all the particles in the cell are of type A and intense blue means that they are all of type B. Intermediate colors indicate that the cell contains both types of particles (white correspond to $50\%$ of each type of particle). The curved interface represent the isosurface $\Delta \rho({\textbf x}) = 0$ as obtained from the linear interpolation of the density on the grid points. The second interface due to the periodic boundary conditions on the long edge of the simulation box is not shown.}
\label{fig:dRho-FieldAndIso}
\end{figure}

In Fig. (\ref{fig:dRho-FieldAndIso}) is reported the $\Delta \rho({\textbf x}, t)$ field on the points $\{{\textbf x_\alpha}\}_{\alpha=1,m}$ together with the isosurface $\Delta \rho({\textbf x}) = 0$ obtained as a linear interpolation of the value of the field on the grid points. This figure shows that the interface is rather sharp, involving typically one or two shells along the direction orthogonal to it. 

From each of these initial conditions we start 25,000 timestep long unrestrained MD simulations along which we compute the microscopic observables of interest. By averaging over the (40) initial conditions we get $\rho^A({\textbf x}, t)$ and ${\textbf v}^A({\textbf x}, t)$.  

\subsection{Results and discussion.}
\label{Sec:ResultsAndDiscussion}
In this section we present our results obtained with the restrained MD method and compare them with those obtained by computing the relevant fields along one unrestrained trajectory started from a configuration sampled from the restrained dynamics. This second type of simulation, often combined with a ``local time average'' (i.e. averaging over a small time-window centered at the current time), are used to study hydrodynamical phenomena  by atomistic simulation. We show that the fields obtained from this latter type of simulation violate some of the properties of the hydrodynamical fields associated to the process under investigation while our D-NEMD approach does not. 

\begin{figure}[h]
\includegraphics[width=1\textwidth]{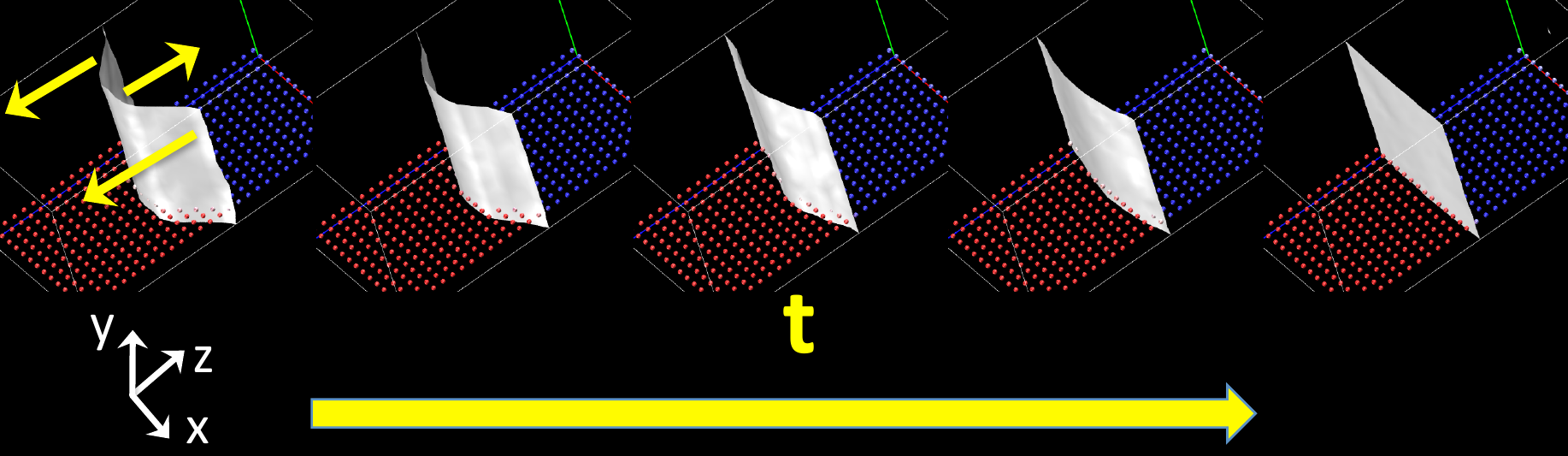}
\caption{Snapshots of the interface $S(t)$. The arrows on the first  snapshot show the direction of evolution of the interface: the center and the extreme of the interface move in opposite directions. The  field $\Delta \rho({\textbf x},t)$ is also shown on one $\{xz\}$ plane by adopting the same colorcoding of Fig. \ref{fig:interfaceTimeline}}
\label{fig:interfaceTimeline}
\end{figure}

Let us start by analyzing the surface $S(t) = \left \{{\textbf x}: \Delta \rho({\textbf x},t) = 0\right \}$. In Fig. (\ref{fig:interfaceTimeline}) it is shown a series of snapshots of the $S(t)$ surface. First of all we remark that all along its evolution the surface satisfies the symmetry of the problem, i.e. it is symmetric with respect the $\{yz\}$ reflection plane passing by the middle of the simulation box and it is translationally invariant along the $y$ direction. The tiny bumps on the $S(t)$ surface are due to the limited number of initial conditions (40) used for computing $\Delta \rho({\textbf x},t)$. In the limit of an infinite number of such initial conditions the surface would be completely smooth, as expected and predicted by classical hydrodynamics. The relaxation from the initial curved surface to the final flat surface takes approximately $20,000$ timesteps which, if we chose for the Lennard-Jones parameters the values $\sigma=3.405$~\AA\ and $\epsilon=0.01032$~$eV$ (suitable for modeling Ar) and an amplitude of the initial interface ${\mathcal A} = 50$~\AA, corresponds to a maximum value of the field ${\textbf v}({\textbf x}) $ of $\sim 80~m/s$. 

\begin{figure}[h]
\includegraphics[height=0.6\textheight]{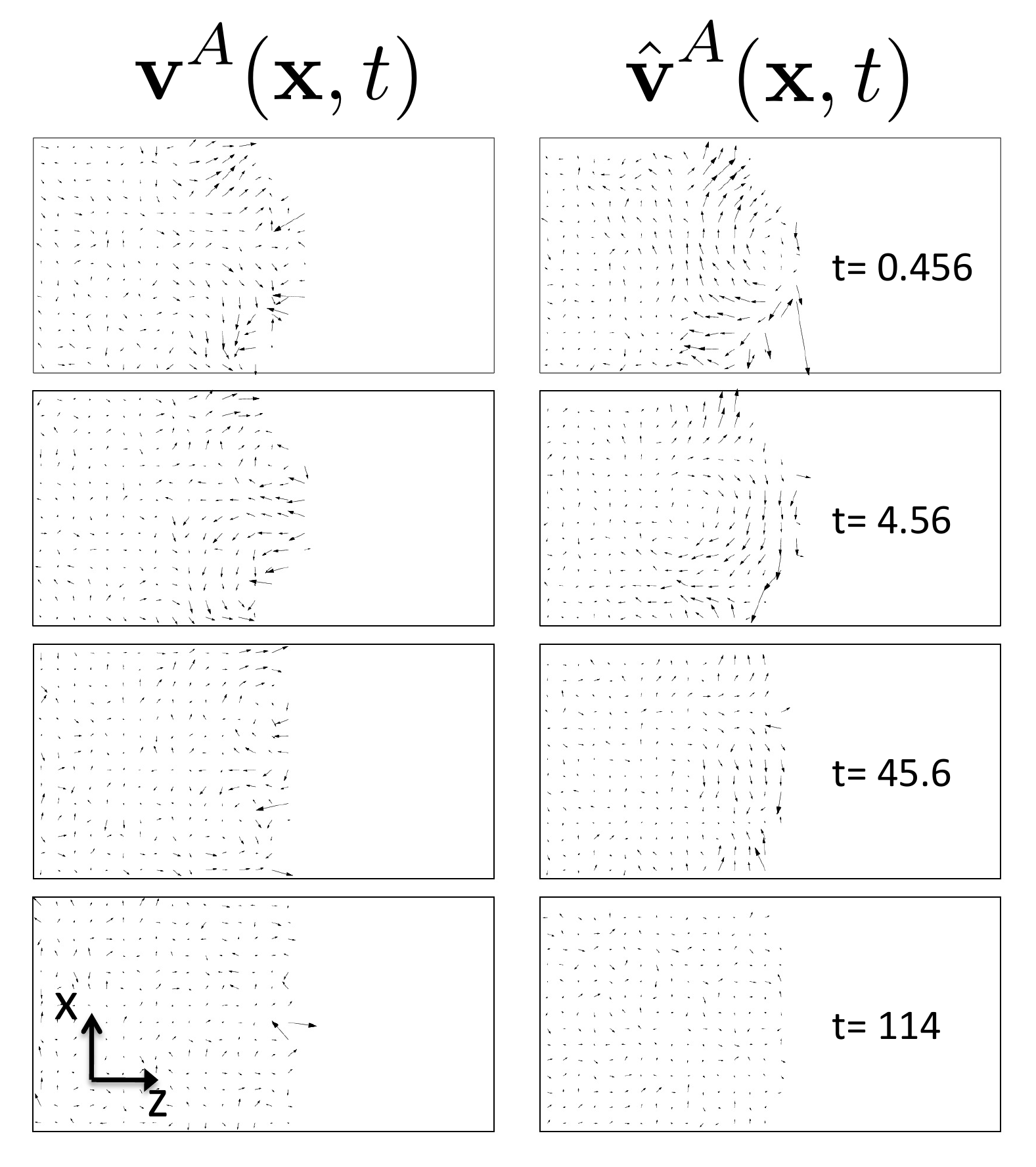}
\caption{Snapshots of the ${\textbf v}^A({\textbf x}, t)$ (left, see Eq. (\ref{eq:fieldVelocity})) and ${\hat {\textbf v}}^A({\textbf x}, t)$ (right, see Eq. (\ref{eq:velField-Single})) fields on the grid points belonging to one $\{xz\}$ plane at various times. }
\label{fig:velField}
\end{figure}

We now move to the analysis of the velocity field, focusing on the velocity field of only the chemical species A:

\begin{eqnarray}
\label{eq:velField}
{{\textbf v}}^A({\textbf x}, t) &=& \\
&&{Ê \sum_{\Gamma_0} \sum_{i=1}^{N_A} {\textbf p}_i(\Gamma_0) \Pi_{\chi=1}^3 \left [ erf({\overline x}_{\alpha,\chi} -  r_{i,\chi}(\Gamma_0), \sigma) - erf({\underline x}_{\alpha,\chi} -  r_{i,\chi}(\Gamma_0), \sigma) \right ] \over Ê \sum_{\Gamma_0} \sum_{i=1}^{N_A} \Pi_{\chi=1}^3 \mu_i \left [ erf({\overline x}_{\alpha,\chi} -  r_{i,\chi}(\Gamma_0), \sigma) - erf({\underline x}_{\alpha,\chi} -  r_{i,\chi}(\Gamma_0), \sigma) \right ] } \nonumber
\end{eqnarray}

\noindent  where the sum $\sum_{i=1}^{N_A}$ runs only over the atoms of type A, and it is implicitly assumed that ${\textbf p}_i$ and ${\textbf r}_i$ are taken at the time $t$ starting from the initial condition $\Gamma_0$ (see Eqs. (\ref{eq:field-Discrete-Smooth}) and (\ref{eq:fieldVelocity})). The sum $\sum_{\Gamma_0}$ run over the initial conditions along the restrained MD.
We consider the field  ${{\textbf v}}^A({\textbf x}, t)$ as, due to the conservation of the total momentum and the fact that the initial total momentum was set to zero, the total field, i.e. those including A and B specie, is, on average, zero. In the left column of Fig. \ref{fig:velField} is shown the velocity field ${\textbf v}^A({\textbf x}, t)$ at various times. This field is computed only on the grid points corresponding to cells that contains at least one particle of type A. This fact makes the field ``noisy'' (large values of the field rapidly changing orientation) close to the A/B interface, where the cells contain less A particles, and therefore the average of the atomic velocities over the particles in the cell (see Eqs. (\ref{eq:field-Discrete}) and (\ref{eq:field-Discrete-Smooth})) is less effective in smoothing the field. This effect is reduced by making larger the number of initial configurations used to perform the ensemble average over the initial conditional PDF.
As a first remark, it is worth to mention that relatively few cell layers nearby the interface are involved in the relaxation process. In fact,  already $5-10$ cells far from the interface the velocity field is essentially zero at any time during the relaxation. Coming to the hydrodynamical process producing the relaxation of the interface, we notice that initially (see panel 1 of Fig. \ref{fig:velField}) the velocity field at the top of the interface is pointing downward while at bottom it is pointing upward. After some time this field stabilizes into a double symmetric roll, one rotating clockwise and the other one rotating counter clockwise, both starting at the top of the interface and ending at its bottom (see panel 2 of Fig. \ref{fig:velField}). Overall, this velocity field produce the phenomenon of pushing up the side of the interface and pulling down the center as shown in Fig. \ref{fig:interfaceTimeline}. The relaxation of the interface follows this mechanism almost till the end of the process. In fact, in panel 3 of Fig. \ref{fig:velField} we see that still after 45.6 LJ time units ($10^4$ timesteps), when the interface is almost flat, the double roll is still present. Eventually, after 114 units of time ($2.5 \times 10^5$ timesteps) the  interface is completely flat and the field is null everywhere (panel 4).

It is very interesting to compare the ${\textbf v}^A({\textbf x}, t)$ field as obtained from the D-NEMD simulation with the {instantaneous} field ${\hat {\textbf v}}^A({\textbf x}, t)$ defined as

\begin{equation}
\label{eq:velField-Single}
{\hat {\textbf v}}^A({\textbf x}, t) = {Ê \sum_{i=1}^{N_A} {\textbf p}_i \Pi_{\chi=1}^3 \left [ erf({\overline x}_{\alpha,\chi} -  r_{i,\chi}, \sigma) - erf({\underline x}_{\alpha,\chi} -  r_{i,\chi}, \sigma) \right ] \over Ê\sum_{i=1}^{N_A} \Pi_{\chi=1}^3 \mu_i \left [ erf({\overline x}_{\alpha,\chi} -  r_{i,\chi}, \sigma) - erf({\underline x}_{\alpha,\chi} -  r_{i,\chi}, \sigma) \right ] }
\end{equation}

\noindent Few snapshots of this field are shown in Fig. \ref{fig:velField}. First of all we notice that the interface relaxation process occurs via the formation of a clockwise roll, which is initially at the top of the interface (panel 1 of Fig. \ref{fig:velField}) and then move toward the bulk (panel 2 of the same figure). This roll is stable all over the duration of the simulation (see panel 3) and eventually disappear when the equilibrium is reached (panel 4). The shape of the field ${\textbf v}^A({\textbf x}, t)$ contrasts with the symmetry imposed on the problem, which implies a $\{yz\}$ mirror plane passing through the middle of the simulation box. This problem cannot be solved by the ``local time average'' that is often used in simulation of hydrodynamical processes by atomistic simulations. In fact, as mentioned above, the clockwise vortex is very stable and a local time average will not restore the proper symmetry expected for this field. This problem illustrates that a proper statistical average is needed in order to compute by atomistic simulation hydrodynamic fields as otherwise a unlucky  choice of the initial conditions can produce unphysical results.


\section{Conclusions.}
\label{Sec:Conclusions}
{In this paper we have reviewed the dynamical approach to non-equilibrium MD. We have shown that using a proper simulation setup it is possible to treat interesting cases in which the initial condition is either a stationary non-equilibrium condition or a constrained equilibrium consistent with the value of a macroscopic scalar or field-like observable. We illustrated the functioning of the method by applying it to two cases: the establishing of convective cells and the relaxation of an interface between two immiscible liquids. We have shown that our method generates rigorous time-dependent non-equilibrium averages,  while the method of local time average, often used to simulate hydrodynamical processes from atomistic simulation, can, sometimes, fail.}

Our conclusion is that the method is ready for challenging applications. Work is in progress in this direction.
 





\begin{theacknowledgments}
S. M. and G. C. acknowledge SFI Grant 08-IN.1-I1869 for the financial support. S.O. acknowledges SimBioMa for the financial support.
Finally, the authors wish to acknowledge the SFI/HEA Irish Centre for High-End Computing (ICHEC) for the provision of computational facilities.

\end{theacknowledgments}



\bibliographystyle{aipproc}   


\end{document}